\documentclass[
    ,final            
  ]{aipproc}

\layoutstyle{6x9}

\begin{document}

\title{Do We Live in a Vanilla Universe? \\
 Theoretical Perspectives on WMAP }

\author{Richard Easther}{
  address={ISCAP, Columbia Astrophysics Lab, Mailcode 5247, 550 W
  120th Street, New York, NY10027 }}

\begin{abstract}
I discuss the theoretical implications of the WMAP results, stressing
WMAP's detection of a correlation between the E-mode polarization and
temperature anisotropies, which provides strong support for the
overall inflationary paradigm.  I point out that almost all
inflationary models have a ``vanilla limit,'' where their parameters
cannot be distinguished from a genuinely de Sitter inflationary phase.
Because its findings are consistent with vanilla inflation, WMAP
cannot exclude entire classes of inflationary models.  Finally, I
summarize hints in the current dataset that the CMB contains relics of
new physics, and the possibility that we can use observational data to
reconstruct the inflaton potential.

\end{abstract}

\maketitle


\section{Introduction}

On February 12, 2003, the Wilkinson Microwave Anisotropy Probe [WMAP]
reported results based on its first year of observations
(e.g.
\cite{Bennett:2003bz,Spergel:2003cb,Hinshaw:2003ex,%
Peiris:2003ff,Kogut:2003et}), and cosmology took a giant step towards
its long promised ``golden age.''  Ref.~\cite{Spergel:2003cb} lists
the values of 22 cosmological parameters determined using the WMAP
data and other recent observational information. Many of these values
are quoted with several significant figures, whereas a decade ago they
were either completely undetermined or had massive uncertainties.

For the theoretical cosmologist, the WMAP results are more tantalizing
than revolutionary. On the one hand, WMAP confirms that there is a
significant contribution from dark energy in the present epoch, puts
tight constraints on the parameter space open to inflation, and
provides strong support to the overall inflationary paradigm. However,
this does not surprise most theoreticians, and nothing in the current
WMAP dataset puts genuinely nontrivial constraints on the physics of
inflation and the very early universe.

To explain further, consider {\em vanilla inflation\/} -- an almost
exactly de Sitter inflationary epoch which lasts long enough to
deliver a primordial universe -- whose measurable parameters all have
their ``default values''.  After vanilla inflation, scalar
perturbations are Gaussian and scale-free, and there is no discernible
contribution from tensor modes or curvature in the present epoch.
Consequently, all parameters which constrain the primordial universe
would be measured as upper bounds, rather than definite values.

Almost all inflationary models have parameters which can be tuned to
provide a vanilla limit.  In some cases these tunings may appear so
unnatural, and one may want to exclude the model on aesthetic grounds.
For instance, the perturbation spectrum associated with a $\phi^n$
potential becomes more strongly scale dependent as $n$ increases, and
one may argue that $n$ should be an even, positive number.  The
vanilla limit of this model is $n \rightarrow 0$, so if $n \ge 2$ is
experimentally excluded, the model becomes less attractive.  However,
this prejudice is far less compelling than the observation of an
unambiguously blue spectrum (one with more power on short scales than
on long scales): in this case {\em all\/} positive values of $n$ are
excluded. The one type of inflation which is hard to tune is de Sitter
inflation, where the inflaton is trapped in a local minimum of the
potential and does not evolve at all. However, in this case the
spectrum is precisely scale invariant, which is the vanilla result.

Vanilla inflation has an ambiguous position in theoretical cosmology.
It has the most ``natural'' set of parameter values but offers the
smallest leverage for discriminating between different realisations of
inflation, and thus the least insight into the early universe.
Consequently, theoreticians tend to seize on any hints that the
inflationary epoch contains non-vanilla flavorings, since these
significantly constrain the inflationary parameter space and,
in extreme cases, challenge the overall paradigm.  The first year WMAP
dataset has two tantalizing features: the apparent lack of power at
long wavelengths, and the suggestion that the scalar spectral index
itself is a function of the perturbation's wavelength.

\section{The CMB and Inflation}

Almost any possible inflationary epoch can be described in terms of a
scalar field $\phi$ moving in a potential, $V(\phi)$.\footnote{I am
restricting myself to single field models, but all of the statements
below have an analogous (although often weaker) form for multi-field
models.}  From the Einstein field equations and the energy momentum
tensor for a minimally coupled scalar field once can deduce
\begin{equation}
H^2 = \left({\dot a \over a}\right)^2 = {8 \pi \over 3 m_{\rm Pl}^2}
 \left[{1 \over 2} \dot\phi^2 + V(\phi) \right],
\label{eqbackgroundequation1}
\end{equation}
where $a(t)$ is the spacetime scale factor, and a dot denotes
differentiation with respect to time.  We are implicitly assuming
a spatially flat, homogeneous and isotropic universe where the
inflaton field is the only contribution to the energy-momentum
tensor. The motion of the field is given by
\begin{equation}
\ddot \phi + 3 H \dot \phi + V'(\phi) = 0,\label{eqequationofmotion}
\end{equation}
where the dash notes differentiation with respect to $\phi$. 

Guth's original paper on inflation \cite{Guth:1980zm} addressed
problems associated with the dynamics of the cosmological
background. Inflation can be implemented in a multitude of different
ways, all of which solve the cosmological problems addressed by Guth.
Consequently, inflationary model builders do not focus directly on the
expansion history of the universe.  In addition to the zero-order
dynamics needed to set the stage for a hot big-bang universe,
inflation also predicts the first order perturbations about this
background solution.  These perturbations determine both the
clustering properties of galaxies, and the anisotropies in the
microwave background.

Inflation produces primordial perturbations by magnifying quantum
fluctuations until their wavelength is equal to or larger than the
present size of the observable universe.  The properties of these
fluctuations differ markedly between different inflationary models and
specific inflationary scenarios can thus be distinguished from one
another and tested via their perturbation spectra.  Consequently,
putting tight experimental constraints on the perturbation spectrum is
of prime importance, since the theoretical cosmologist can use this
data to eliminate specific inflationary models. However, almost all
inflationary models have a vanilla limit, so as long as vanilla
inflation remains consistent with the observational data we cannot
exclude entire classes of models.

Given the functional form of the potential and fairly mild assumptions
about the dynamics of inflation, we express the perturbation spectra
as a function of the potential and its derivatives. A general
perturbation to the background $g_{\mu\nu}$ is a symmetric tensor,
$h_{\mu\nu}$, where the perturbed spacetime is $g_{\mu\nu} +
h_{\mu\nu}$ \cite{Mukhanov:1990me}. We decompose $h$ into scalar,
vector and anti-symmetric tensor components, where the decomposition
reflects the transformation properties of the different pieces under
(small) transformations. Cosmologically, we need consider only the
scalar and tensor modes, as the vector modes decay with time.  The
scalar modes are associated with a gravitational potential and are the
source of density fluctuations in the universe. The tensor modes are
effectively gravity waves (and are often referred to as such) and do
not contribute to the formation of structure in the universe, but do
contribute to the microwave background anisotropies, especially at
large angular scales.

The perturbations are described in terms of their power spectra, 
\begin{equation}
P_S^{1/2} \sim k^{n_S-1}, \quad
P_T^{1/2} \sim k^{n_T} \label{nsnt}
\end{equation}
where $k$ is the comoving wavenumber of the perturbation. If $n_S=1$
or $n_T =0$ the amount of power in each mode is independent of $k$ and
the resulting spectra are {\em scale invariant\/}.\footnote{ The
differing definitions of $P_S$ and $P_T$ are an historical anomaly.}
We know that the underlying spectra must be roughly scale invariant,
but the question is whether the difference between $n_S$ and $n_T$
from their ``natural'' values of 1 and 0 is detectable
observationally.

Using the slow roll approximation, we can write $n_S$ and $n_T$ in
terms of derivatives of the potential \cite{Lidsey:1995np},
\begin{equation}
n_S = 1 + 2 \eta - 4 \epsilon, \qquad n_T = -2 \epsilon 
\end{equation}
where 
\begin{equation}
\epsilon =  \frac{m_p^2}{16 \pi} \left(\frac{V'}{V}\right)^2\, , \quad 
\eta = \frac{m_p^2}{8 \pi} \left[ \frac{V'}{V} -
 \frac{1}{2} \left(\frac{V'}{V}\right)^2 \right].
\end{equation}

\begin{figure}[tbp]
\includegraphics[height=6cm]{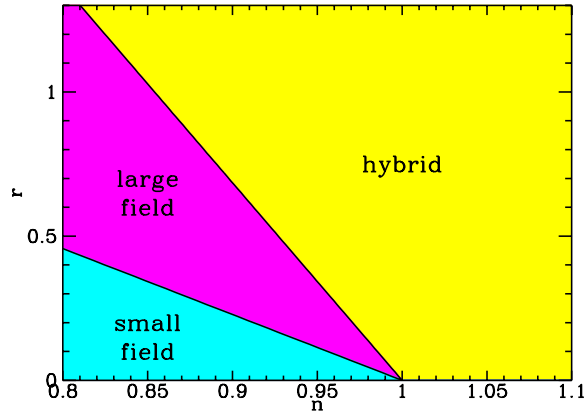}
\caption{The three classes of single field inflationary models are
displayed.  The hybrid models have $\eta>\epsilon>0$, and the small
field models have $\eta<0$. Vanilla inflation corresponds to $r=0$ and
$n=1$, the point at which the three wedges intersect. (Plot from
\cite{Kinney:1998md}.)}
\end{figure}

Finally, the amplitudes of the two spectra must obey a consistency
condition, 
\begin{equation}
\frac{P_T^{1/2}}{P_S^{1/2}} = 16 \epsilon 
\end{equation}
where the numerical coefficient is, to some extent, a matter of
definition.  We can divide inflationary models into three general
classes, summarized in Figure~1 \cite{Kinney:1998md}. In the hybrid
case, the $V''/V \gg (V'/V)^2$, and the field is evolving towards a
local minimum of the potential.  Conversely, if we find $n_S<1$ and no
detectable tensor contribution, $V''<0$ and $V'\sim 0$, and we have
{\em small field inflation\/}.  Finally, if we observe a significant
tensor component, we must have $V'/V \ne 0$. From the equation of
motion for $\phi$, we see that $\dot{\phi} \ne 0$, and the total
evolution of $\phi$ during inflation can be substantial -- leading to
the moniker {\em large field inflation\/}.

While a non-zero scalar spectrum is needed to provide the primordial
density fluctuations that seed the formation of structure, a
primordial tensor spectrum is optional. We can show that the
$P_T^{1/2}$ is proproportional to the value of $H$ (and thus the
square root of the energy density) during inflation and, unless $H$ is
GUT scale or above, the tensor signal will most likely be forever
undetectable. Since a detectable tensor signal is produced by a
limited range of inflationary models the vanilla prediction is that
the CMB contains no detectable contribution from tensors.  However, if
we {\em do\/} observe a primordial tensor spectrum, then we can
immediately deduce the energy scale at which inflation occurred.
Moreover, if we can measure both its amplitude and index, we are in
the pleasant position of having four observable quantities (the
amplitudes and indices of the tensor and scalar spectrum) which are
specified in terms of three parameters. This leads to a ``consistency
condition'' which must be satisfied (to first order in slow-roll) by
all single field inflationary models.

\section{The WMAP Results: Summary}

Conceptually, the WMAP mission is very simple: over a period of
several years, it makes repeated observations of the microwave sky,
and is sensitive to both temperature and polarization. It observes in
five frequency bands, since the main foreground contaminants scale
differently with frequency from the underlying black body of the CMB.

\begin{figure}[tbp]
\includegraphics[angle=90,width=0.6\textwidth]{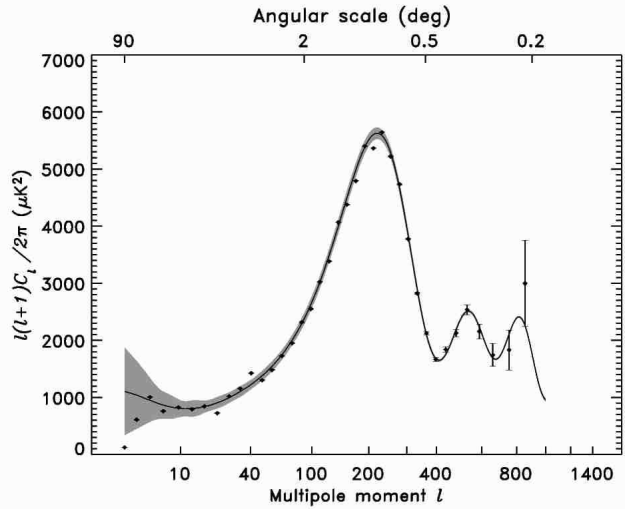}
\caption{The WMAP power spectrum, where the shaded region represents
the expected cosmic variance about the best fit spectrum
\cite{Hinshaw:2003ex}.}
\end{figure}
 
Information can be extracted from the maps directly (for instance, the
topology of isotemperature contours is a function of the underlying
cosmological model \cite{Colley:2003sp}), but the maps are frequently
distilled into a power spectrum by expanding them in spherical
harmonics, and the WMAP power spectrum is shown in Figure 2. Different
theoretical models of the early universe predict different scalar and
tensor spectra, but the CMB also depends on parameters such as the
cosmological constant and the Hubble constant via their influence on
the evolution of the perturbations as they propagate in an expanding
universe. This is crucial, since it turns the CMB into a tool for
estimating a wide range of cosmological parameters which, taken
together, put tight constraints on the composition and history of our
universe. 

Given a set of parameters, the theoretical spectrum is estimated using
a tool such as CMBFAST. Armed with high quality CMB data (and data
from other sources) one can find the ``best fit'' model by varying the
parameters until the underlying spectrum is matched as accurately as
possible. Spergel {\it et al.\/} \cite{Spergel:2003cb} describes this
process.  The WMAP team concludes that the age of the universe is
$13.7\pm 0.2$Gyr, the total mass-energy of the universe is
$\Omega_{tot}= 1.02 \pm 0.02$ (where a value of unity corresponds to a
universe with no spatial curvature), the parameterized Hubble constant
$h=0.71^{+0.04}_{-0.03}$, and baryon density (as a fraction of the
total energy density) $\Omega_b h^2 = 0.0224 \pm 0.0009$.  Little more
than a decade ago, the value of all of these numbers was the subject
of significant controversy, and yet here they are all quoted to two or
three significant figures.

\section{Polarization}

Since the first detection of the CMB anisotropies, attention has
focussed on the temperature variations.  The polarization also varies
from point to point, but this anisotropy is intrinsically smaller and
more difficult to measure.  WMAP is the first full-sky mission to
return a non-zero polarization measurement
\cite{Kogut:2003et}.\footnote{The polarization itself was first detected 
by the DASI mission \cite{Kovac:2002fg}.}

\begin{figure}[tbp]
\includegraphics[angle=0,width=0.6\textwidth]{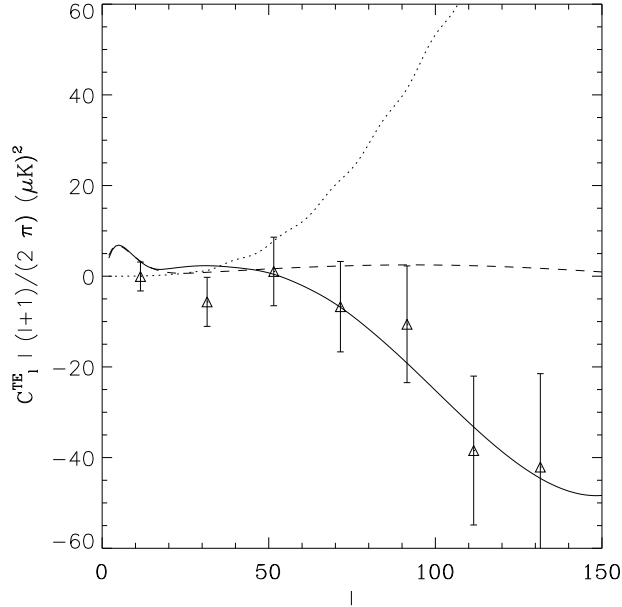}
\caption{The cross-correlation between the E-mode polarization and
temperature anisotropies in the CMB. The solid line represents the
inflationary prediction, while the defect prediction is shown by the
dotted line \cite{Hinshaw:2003ex}.}
\end{figure}

In a universe where inflation occurs, the primordial perturbations may
be correlated on scales far larger than the present size of the
observable universe, and were definitely correlated on super-horizon
scales when the microwave background photons decoupled from the rest
of the universe some 300,000 years after the big bang.  While other
mechanisms for generating the primordial density perturbations have
been explored (particularly ``defect models'' where objects such as
cosmic strings generate perturbations as they move through the
universe), these do not produce super-horizon
correlations.\footnote{The ekpyrotic \cite{Khoury:2001wf} and pre big
bang \cite{Gasperini:1992em} scenarios both produce super-horizon
correlations, but in a universe which is initially contracting. While
the proponents of these models typically define them in contrast to
inflation, they resemble inflation more than they resemble defect
models of structure formation. In particular, like inflation, they
possess an era in which $\ddot{a}>0$ and modes with fixed comoving
size are continually leaving the horizon.} A frequent critique of
inflation is that the overall paradigm makes no generic predictions,
and that there is no one property that all inflationary models share.
However, a universe with super-horizon perturbations at decoupling has
a characteristic correlation between the temperature anisotropies and
the E-mode polarization signal. WMAP was observed this correlation,
and found that it closely matches the inflationary prediction.

It is worth pausing to reflect what a significant achievement this is:
from my perspective the observed correlation between temperature and
E-mode polarization is most important single result in the WMAP
dataset.  Prior to inflation, no-one had predicted super-horizon
correlations, so this observation is a stunning verification of one of
the key features of inflation.  Moreover, if the data had not
confirmed the inflationary prediction, almost all models of inflation
would have been ruled out in a single stroke.\footnote{Inflation may
still have occurred, but it could not have produced the observed
primordial perturbations.}  Consequently, this is a key test of the
inflationary paradigm. The only downside (and perhaps the reason why
more is not being said about it) is that because all inflationary
models make this prediction, it only eliminates models such as defect
scenarios which have already fallen from favor in the theoretical
community.  However, I suspect that this observation will be regarded
as one of the lasting achievements of WMAP, and that observations of
the $\langle TE \rangle$ cross-correlation will be followed carefully
as the data improves.

\section{Surprises: Non-Vanilla Features in WMAP?}

The two obvious peculiarities in the WMAP dataset are that the spectrum
has anomalously low power on small scales, and that the fit to the
data improves if the scalar index is allowed to be scale-dependent
(that is, $n_s$ is a function of $k$).  Either of these results {\em
could\/} provide a dramatic non-vanilla flavor to the early universe.
However, in both cases their physical significance is hard to quantify
and await both further data and theoretical analysis.

\subsection{Low Quadrupole}

The low quadrupole is visible in the power spectrum of Figure~2, as
the first data points lie well below the best-fit spectrum.  Due to
cosmic variance\footnote{The $C_l$ values plotted in Figure~2
represent averages over the $2l+1$ $m$ values of the corresponding
$Y_{lm}$. Roughly speaking, this average will have a sampling
uncertainty of $1/\sqrt{2l+1}$ -- but since we can observe only one
sky we cannot reduce this uncertainty by gathering more data. This
intrinsic uncertainty is the cosmic variance.}, we don't expect an
exact match between theory and experiment at small values of $l$, but
the discrepancy is, on the face of it, unexpectedly large.

While it is clear that the observed CMB sky has less power at large
wavelengths (low $k$) than that suggested by the ``best fit'' $\Lambda$CDM
model, it is not clear is whether this is something we {\em need\/} to
explain, since the result could be produced by cosmic variance alone.
Spergel {\it et al.\/} \cite{Spergel:2003cb} generated multiple
realizations of the microwave background with the parameters estimated
by WMAP, and found that the probability that the low power at small
$l$ is due to cosmic variance is $1.5 \times 10^{-3}$.  However, other
authors (\cite{Efstathiou:2003wr}, for example) argue that this
calculation underestimates cosmic variance, and that the discrepancy
is not large enough to be a signal of new physics.

If this result is significant, the possible modifications to the
standard paradigm that would suppress power at large scales take a
variety of forms. For example, a comparatively conservate approach
is provided by Contaldi {\it et al.\/} \cite{Contaldi:2003zv}, who
look at inflationary models which are tuned to suppress $P_S$ for
values of $k$ which dominate the low $l$ terms in the CMB spectrum.
Carefully tuned inflationary models violate the spirit of the
inflationary paradigm, but they are less radical departures from the
standard cosmology than (for instance) advocating a toroidal universe
with a ``cell size'' that is smaller than the current size of the
visible universe \cite{deOliveira-Costa:2003pu}, or that the universe
possesses detectable (and positive) spatial curvature
\cite{Efstathiou:2003hk}, both of which would tend to cut off the
power spectrum at long wavelengths.

Simply measuring the microwave sky more accurately will not reduce the
cosmic variance. However, it is possible that some of lack of power at
low $l$ could be explained by an over-aggressive foreground
subtraction, and this is amenable to testing and improvement.
Conversely, if the suppression of power at low $l$ is a real effect,
evidence for it will appear in other places.  For example, Kesden {\it
et al.\/} \cite{Kesden:2003zm} show that the shear produced by
gravitational inhomogeneities, which distorts the correlation between
temperature and polarization anisotropies, would be measurably
different in a universe where the spectrum lacked power at small
values of $k$, and this can be tested by future experiments.

\subsection{Running Index}

The low quadrupole seen by WMAP is suggestive of unsuspected physics
that plays a role at large angular scales. However, WMAP also hints
that the standard assumption of an underlying spectrum described by a
constant index $n_S$ may be too simplistic \cite{Peiris:2003ff}. In
this case, the principal evidence is found in CMB data at small
angular scales. This is currently dominated by observational
uncertainty, rather than cosmic variance. In fact, the first-year WMAP
dataset alone does cover a large enough range of $l$-values to provide
any evidence for a running ($k$ dependent) $n_S$. The evidence for
running (between the 1 and 2 $\sigma$ level) appears when the WMAP
power spectrum is combined with that derived from galaxy surveys and
Lyman-$\alpha$ forest data, both of which provide data on the
primordial spectrum at smaller scales than is possible with the CMB
alone \cite{Peiris:2003ff}.

If the running index is confirmed, it will put tight constraints on
inflation.  The slow roll expansion for the spectral index given by
equation~(\ref{nsnt}) can be extended to beyond the leading order
result given here, and $dn_{S}/d\ln{k}$ is dependent on the third
derivative of the potential. However, having $V'''(\phi)$ large enough
to produce a $dn_{S}/d\ln{k}$ observable by WMAP rules out almost all
standard models of inflation.  This is not, in itself, a drawback.
Moreover, it would comprehensively rule out vanilla inflation, which
would be very a welcome development indeed.  Moreover, if the
potential has a number of ``features'' then it may not need to be
carefully tuned in order to ensure that one of these features is found
within the range of $\phi$ covered by the inflaton field as the
cosmologically relevant perturbations are generated
\cite{Adams:2001vc}. Indeed, a potential of this sort was considered
by the WMAP team, and there is weak support in the WMAP dataset (in
combination with other survey information) for this type of feature.

The major caveat about this possible non-vanilla flavoring of the
early universe is simply that the data is inconclusive. The result
hinges on the merger of several datasets, which increases the
complexity of any statistical analysis, and the level of significance
is small enough for it to simply be the result of a statistical
fluctuation. This uncertainty will soon be resolved -- the next year
of WMAP data will significantly improve the sensitivity of the
measurements of the $C_l$ for larger values of $l$, and the SDSS
[Sloan Digital Sky Survey] will supplant the galaxy surveys used by
the WMAP team in their previous papers.  

\section{Reconstructing the Potential}

One of the principal dreams of the theoretical cosmologist is to
reconstruct the underlying physical mechanism of inflation. In
general, this amounts to recovering the functional form of the
potential. Even in a ``golden age'' this inverse problem remains
enormously difficult. Several efforts have been made to develop a
methodology for reconstructing the potential from its Taylor series,
but these appear to be best by observational difficulties
\cite{Lidsey:1995np}. More recently, Easther and Kinney developed
{\it Monte Carlo reconstruction\/}, a stochastic approach to the
problem based on generating a large number of ``trial'' inflationary
models and isolating those for which the observable parameters
coincide with the window in parameter space permitted by the available
data\cite{Easther:2002rw}.  This approach builds on a thorough
understanding of the ``flow equations'' \cite{Kinney:2002qn}, a
consistent expansion of the inflationary dynamics.\footnote{See also
Liddle's recent paper \cite{Liddle:2003py}.} If the permitted window of
parameter space is sufficiently narrow the class of allowed potentials
will be sufficiently well-defined that one can then proceed to
estimate the functional form of the potential. 

The WMAP team used a variant of Monte Carlo reconstruction
\cite{Peiris:2003ff}. This problem has been tackled in more detail by
Kinney {\it et al.\/} \cite{Kinney:2003uw}, who find three classes of
reconstructed potential, corresponding to the three subdivisions of
the ``zoo plot'' shown in Figure~1.  This is of course expected, given
that vanilla inflation remains consistent with the observational data.
However, it is only with the release of WMAP data that the
observational constraints on inflationary theories are tight enough
for this sort of calculation to return any non-trivial limits on the
possible range of potentials which could have driven inflation.

\section{Conclusion}

This paper has given a quick overview of theoretical cosmologists'
response to the WMAP data.  WMAP confirms what we already believed we
knew -- that the perturbations are correlated on super-horizon scales
as a result of an inflation(like) mechanism, thanks to the observed
anti-correlation between the temperature anisotropies and the E-mode
polarization signal.  However, at present the observational evidence
is not tight enough to rule out whole classes of inflationary
models. Crucially, vanilla inflation -- which is a limit of almost
all inflationary models -- remains viable. Since vanilla inflation is
allowed, it follows that no classes of model can be excluded, even if
certain parameter values can be ruled out within each class.  I have
reviewed the two hints in the WMAP data for a non-vanilla universe --
the low quadrupole and the possible running scalar index -- and
sketched how these effects could change our understanding of the early
universe, and how future data is likely to constrain them more
closely. In conclusion, though, it is clear that WMAP marks a profound
change in the theoretical debate, and that the ``golden age'' of
cosmology is upon us.

\section*{Acknowledgements}

I thank my Columbia colleagues Ted Baltz, Brian Greene and Will Kinney
for many useful conversations which helped me form the viewpoints
expressed here, and I thank Hiranya Peiris for several useful
discussions about the WMAP results and their interpretation.

\end{document}